\documentclass[a4paper]{toledoStyle}
\usepackage{epsfig}

\begin{document}

\title{N-Body Simulations of Open, Self-Gravitating Systems}

\author{Daniel Huber} 

\institute{Geneva Observatory, CH-1290 Sauverny, Switzerland}

\maketitle 

\begin{abstract}
Astrophysical systems differ often in two points from classical
thermodynamical systems: 1.) They are open and 2.) gravity
is a dominant factor. Both modifies the homogeneous
equilibrium structure, known from classical thermodynamics.
In order to study the consequence for structure formation
in astrophysical systems, we carry out N-body simulations
of self-gravitating systems, subjected to an energy-flow.
The simulations show that physically realistic, time-dependent
boundary conditions can maintain a molecular cloud in a statistically
steady state, out of thermodynamic equilibrium.

Moreover we perform some simple ``gravo-thermal'' N-body experiments
and compare them with theoretical results. We find negative specific
heat in an energy range predicted by \cite*{authorf:Follana99}.

\end{abstract}

\section{Introduction}

An energy-flow through a system is related to an entropy-flow. If the 
entropy-flow leaving the system is larger than those entering, then
the system evacuates it's internal, by irreversible processes produced
entropy to the outer world. Consequently order is created inside the 
system, which diverges from a classical thermodynamical equilibrium in
a closed system. Far from equilibrium, the system is no longer
characterized by an extremum principle, thus losing it's stability. 
Therefore perturbations can lead to long range order, through which
the system acts as a whole. Such a behavior is well known in
laboratory hydrodynamics and chemistry. The underlying concepts such 
as ``{\it dissipative structures}'' and ``{\it self-organization}'' were
extensively studied (see e.g. \cite{authorf:Nicolis77}). But despite
of their popularity they are until now only little studied in the 
context of non-equilibrium structures in self-gravitating
astrophysical systems. Therefore we take up some ideas of these
concepts and build a simple model of an 
{\it open self-gravitating system}. 
With this model we want to check if an energy-flow can maintain a 
self-gravitating system in an statistically stable state, out of 
thermodynamical equilibrium and if gravitation in combination with 
an energy flow can create structures with a higher degree of order.

The simplicity of the model permits, furthermore, to perform some
``gravo-thermal'' experiments whose results can be compared with
theory.  

\section{Why Dissipative N-Body Systems?}

The use of hydrodynamics to simulate self-gravitating, clumpy gas may 
be problematic, due to the following reasons:\vspace{1ex}

\noindent 1.) During recent years observations revealed that molecular 
clouds are structured down to the smallest resolvable scales. These
structures appear fractal. As a consequence density, temperature and
related fields are not everywhere differentiable, rendering the use of 
hydrodynamics problematic.\vspace{1ex}

\noindent 2.) Hydrodynamics is based on the assumption of a local
thermodynamic equilibrium (LTE). But such an equilibrium may not be
established in gravitationally unstable media because the speed of matter
disturbances is comparable to the sound speed
(\cite{authorf:Pfenniger98}).\vspace{1ex}

\noindent 3.) Thermodynamical formalism incorporating gravity yield
results differing from those of classical thermodynamics. An example
of this is the so called gravo-thermal catastrophe. The application
of these formalism lead among other things to spatially inhomogeneous 
equilibrium states. In hydrodynamics gravity is inserted in the Euler
equation as an external force. But the thermodynamical variables are
determined by a theory not taking gravity into account. Because
gravity can change thermodynamical behavior, it is not a priori 
clear that hydrodynamical methods can simulate self-gravitating gas
correctly.

Bearing in mind these considerations we try an other approach in order
to model self-gravitating interstellar gas. That is, we
use dissipative particles, representing dense cloud fragments, 
to simulate cosmic gas (\cite{authorf:Huber01a}). 

Furthermore, with such a realization we can check some thermodynamic 
results of self-gravitating systems with softened potentials.

\section{Model}
\subsection{Principle}

To prevent gravitationally unbound particles from dissolution, the
particles are confined in a spherical potential well. This prevents
a matter flow. However, our system is subjected to an energy flow.
This flow is maintained by energy injection (heating) and dissipation.
The energy is injected due to time and space dependent potential
perturbations. The dissipation is realized by a local or global
cooling scheme.

Among others we apply a particle potential becoming repulsive on the
softening length scale. With such a potential we want to prevent the
formation of very dense particle agglomerates below the resolution
scale. The repulsive force is analogous to the molecular van der
Waals force.

\subsection{Self-Gravity and Repulsion}

\begin{figure}
\centerline{
\epsfig{file=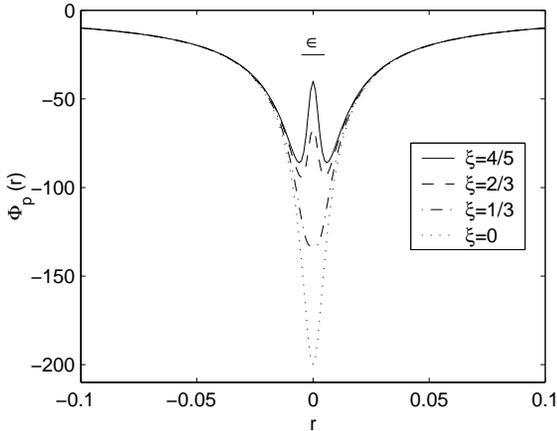,angle=90,width=7.5cm}}
\caption{The potential as a function of $r$ in units of
  $Gm/R$, where $m$ is the particle mass. For $\xi>1/3$ the potential
  is repulsive on the softening length scale.}
  \label{authorf_fig:huberd2_fig1}
\end{figure}

The gravitational forces are computed on the Gravitor Beowulf 
Cluster\footnote{http://obswww.unige.ch/\symbol{126}pfennige/gravitor/gravitor.html}
with a parallel tree code, based on the \cite*{authorf:Barnes86}
tree algorithm. The particle potential is:
\begin{equation}
\label{equpot}
\Phi_{\rm p} (r)=-\frac{Gm}{\sqrt{r^2+\epsilon^2}}\left
      (1-\xi\frac{\epsilon^2}{(r^2+\epsilon^2)}\right)\;,
\end{equation}
where $m$ is the particle mass, $\epsilon$ is the softening length 
and $r$ is the distance from the particle
center. The parameter $\xi$ determines the deviation from the
Plummer potential $\Phi_{\rm Pl}$.
The potential is repulsive 
for $\xi>1/3$ in the range
$|r|<\epsilon\sqrt(3\xi-1)$ (see Figure~\ref{authorf_fig:huberd2_fig1}).

\subsection{Confinement}

The confinement potential, preventing the dissolution of
gravitationally unbound particles has the form,
\begin{equation}
\Phi_{\rm conf} \propto R^{16}\;,
\end{equation}
where $R$ is the distance from the center of the system.

\subsection{Perturbation (Heating Scheme)}

\begin{figure}
\epsfig{file=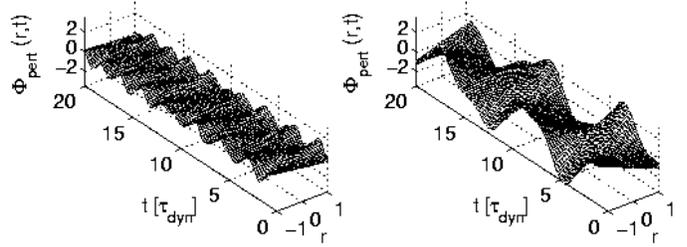,angle=90,width=\hsize}
\caption{Perturbation potentials as a function of time and
  1D-space. The space coordinate r is determined through
  $r^2=x^2+y^2+z^2$ and $x=y=z$. The time is indicated in units of 
  dynamical times. We show two perturbation potentials 
  with different frequency and amplitude. Full resolution version of
  this figure is
  available at: http://obswww.unige.ch/Preprints/dyn\_art.html}
  \label{authorf_fig:huberd2_fig2}
\end{figure}

The energy injection is due to time-dependent boundary conditions or
more precise due to a perturbation potential. The perturbation
potential is shown in Figure~\ref{authorf_fig:huberd2_fig2}. 
The amplitude and the frequency can be 
controlled by parameters. 
If our system represents a molecular cloud then potential
perturbations can be due to star clusters, clouds or other high mass 
objects passing in the vicinity. Indeed such stochastic encounters
must be quite frequent in galactic disks and we assume,
\begin{equation}
\frac{1}{f_{\rm pert}} \ga \tau_{\rm dyn}\;, 
\end{equation}
where $f_{\rm pert}$ is the mean frequency of the encounters.
Thus these encounters can provide a continuous low frequency 
energy injection on large scales.

\subsection{Energy Dissipation}

In order to maintain an energy-flow we must dissipate somehow the
energy injected at large scales. We use two different dissipation
schemes. 

\subsubsection*{1. Local Dissipation}

A particle dissipates energy during an ``inelastic encounter'' with an
other particle. Thus we add friction forces, depending on the 
relative particle velocities $\vec{V}$ and positions $\vec{r}$. 
The friction forces are:
\begin{equation}
F_i \propto \left\{ \begin{array}{rcl}
         \frac{V \cos\vartheta}{r}e_i&:&r>\lambda\epsilon\\
             0 &:& r<\lambda\epsilon
            \end{array}\right.\;,
\end{equation}
where $Vr\cos\vartheta = \vec{V}\vec{r},\; i=\{x, y, z\}$, $\vec{e}$ 
is the unity vector $\vec{V}/V$ and $\lambda$ is a parameter 
$>1$ ensuring the local nature of the dissipation.

\begin{figure*}[!ht]
\centerline{
\epsfig{file=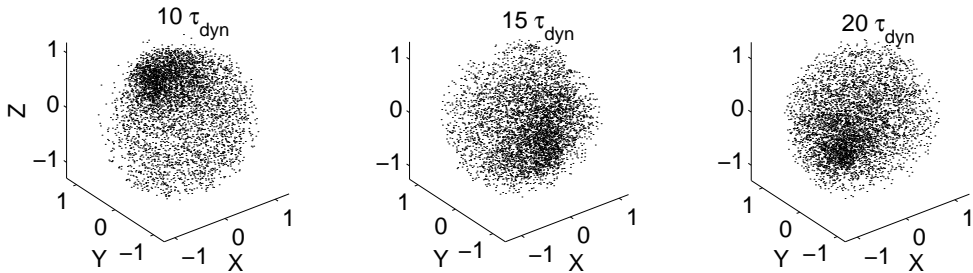,angle=90,width=13cm}}
\caption{A self-gravitating system subjected to an energy-flow. The
  energy-injection compensates the dissipated energy and a
  statistically steady state is reached.}
  \label{authorf_fig:huberd2_fig3}
\end{figure*}

\begin{figure*}[!ht]
\centerline{
\epsfig{file=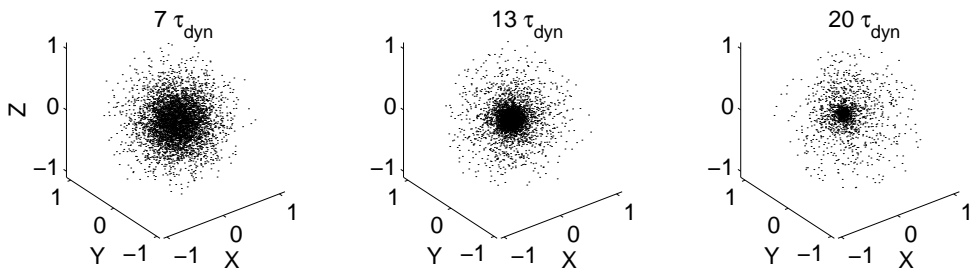,angle=90,width=13cm}}
\caption{The evolution of a self-gravitating system with local
  dissipation, but without heating, i.e., without energy-flow.}
  \label{authorf_fig:huberd2_fig4}
\end{figure*}

\subsubsection*{2. Global Dissipation}

The global dissipation term depends not on the relative velocity $\vec{V}$ 
but on the absolute velocity $\vec{v}$. 
The global friction force, causing an energy dissipation is:
$F_i \propto - v_i$.

\section{Results}

We present here some preliminary results of a paper in preparation
(\cite{authorf:Huber01b}).

\subsection{Self-Gravity and Energy Flow}

With an appropriate tuning of the energy injection we can prevent
the collapse of a self-gravitating dissipative system and maintain it 
in a statistically steady state out of thermodynamic
equilibrium. This is shown in Figure~\ref{authorf_fig:huberd2_fig3}.
The system subjected to an energy flow attains a statistically steady
state after about $10\:\tau_{\rm dyn}$. The evolution of the same
system, but without energy injection, is shown in 
Figure~\ref{authorf_fig:huberd2_fig4}. Both simulations were carried out with
a {\it local dissipation scheme} and 10000 particles, from which 5000
are shown.

\begin{figure}
\centerline{
\epsfig{file=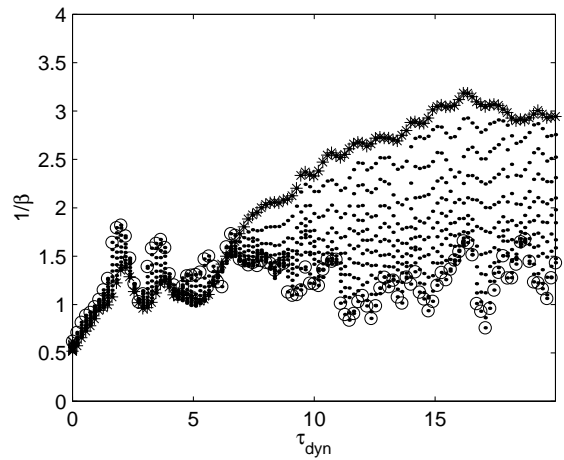,angle=90,width=7.5cm}}
\caption{The evolution of the Lagrangian temperature of the system
  shown in Figure~\ref{authorf_fig:huberd2_fig3}. $\beta$ is the
  inverse dimensionless temperature $(\beta=1/T)$. Each curve depicts
  the temperature of a mass fraction, contained in a sphere. The mass
  spheres are centered at the center of mass and the corresponding
  mass fractions are:
  $\Delta M/M=\{5\%, 10\%, 20\%, \ldots ,80\%, 90\%, 98\%\}$.
  Circles: $\Delta M/M=5\%$. Crosses: $\Delta M/M=98\%$.}
  \label{authorf_fig:huberd2_fig5}
\end{figure}

The system subjected to an energy-flow is not thermalized. A
temperature gradient arises, in such a way that a cooler mass
condensation is embedded in hotter shells 
(see Figure~\ref{authorf_fig:huberd2_fig5}). Such temperature gradients are 
typical for molecular clouds. 

We checked the model behavior in dependence of the different particle
potentials, heating and cooling parameters. Actually we find
statistically stable states, but these states are not endowed with
higher degrees of order.

\section{Gravothermal experiments}

\begin{figure}
\centerline{
\epsfig{file=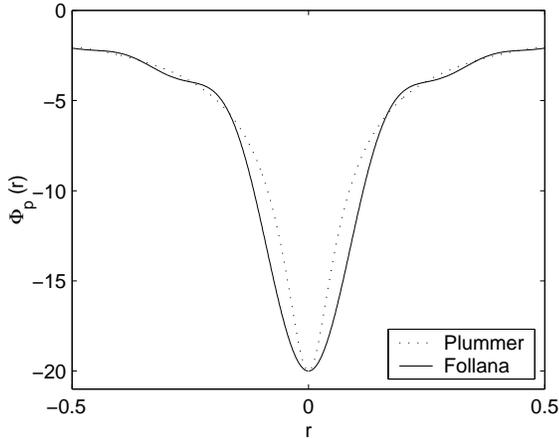,angle=90,width=7.5cm}}
\caption{The Follona $(n=10)$ and the Plummer $(\varepsilon=0.05)$
  potential in units of $Gm^2/R$, where $m$ is the particle mass.}
  \label{authorf_fig:huberd2_fig6}
\end{figure}

If we switch off the heating process and use a global dissipation
scheme we can cool the system and maintain it nearly in thermodynamic
equilibrium. Thus we can perform some simple ``thermodynamic
experiments'' of self-gravitating N-body systems.
Follana \& Laliena examined the thermodynamics of self-gravitating
systems with softened potentials. They achieve a softening by 
truncating to $n$ terms an expansion of the Newtonian potential 
in spherical Bessel functions (we will call this potential 
hereafter Follana potential). This regularization allows the 
calculation of the thermodynamical quantities of self-gravitating 
systems. The form of their potential is similar to a Plummer potential
with a corresponding softening length 
(see Figure~\ref{authorf_fig:huberd2_fig6}). 
Thus we can check their theoretical results by using a Plummer
softened potential.

\begin{figure}
\centerline{
\epsfig{file=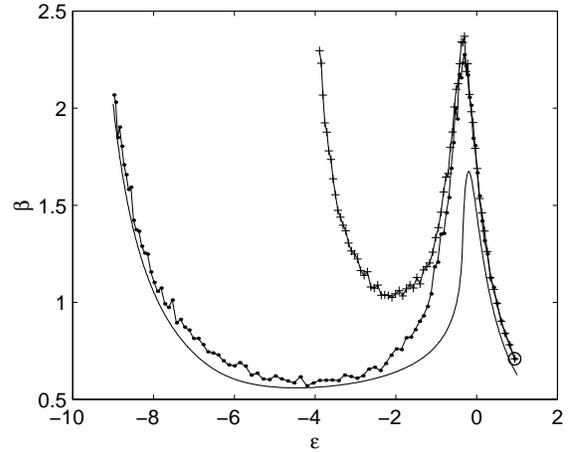,angle=90,width=7.5cm}}
\caption{Inverse temperature $\beta$ versus energy $\varepsilon$
  resulting from models with softened potentials. The solid line
  corresponds to the Follana potential. The other curves depict the
  evolution of the simulated systems. Crosses: Repulsive potential
  $(\xi=2/3)$. Dots: Plummer potential $(\xi=0)$. The circle indicates
  the initial states of the simulations.}
  \label{authorf_fig:huberd2_fig7}
\end{figure}

Because the collapsed phase is probably sensitive to the short
distance form of the potential we expect qualitatively a similar 
behavior only for the Plummer softened potential $(\xi = 0)$ and a 
more different behavior for the potentials with short range 
repulsion $(\xi > 1/3)$. Figure~\ref{authorf_fig:huberd2_fig7} shows the 
evolution of the inverse temperature as a function of the total 
energy\footnote{Temperature and energy are dimensionless, i.e. the units are
$GM^2/R$, where $M$ is the total mass and R is the radius of the
system.} $\varepsilon$
for two simulations with $\xi = 0$ and $\xi = 2/3$, respectively 
as well as the analytical curve calculated by Follana \& Laliena. 
The figure confirms our expectations. We find for the Follana and 
the Plummer potential a negative specific heat in the same energy 
range. The system with a repulsive potential re-enters the zone 
of positive specific heat earlier than those without such a repulsion
below the softening length. After entering the zone of negative 
specific heat a collapsing phase transition takes place, separating 
a high energy homogeneous phase from a collapsed phase. 
The collapsed phase resulting from the N-body simulations shows no 
core-halo structure. Such a structure is only formed due to relaxation
after the cooling has stopped.

\section{Conclusions}

For systems subjected to an energy flow we find:

\begin{itemize}
\item Potential perturbations caused by astrophysical objects passing 
in the vicinity of a self-gravitating system can compensate the 
energy loss due to dissipation, thus prevent the system from 
collapsing and maintain a statistical steady state.

\item The statistical steady state is out of thermodynamic equilibrium 
and consists of a dense cold core moving in a hotter halo.

\end{itemize}

The results of the gravo-thermal experiments are:

\begin{itemize}

\item The range of negative specific heat agrees for a Plummer softening
with the theory of self-gravitating systems with softened potentials.

\item A softened potential incorporating short range repulsive forces
reduces the range of negative specific heat. Systems with such
repulsive forces re-enter the zone of positive heat capacity, thus 
becoming stable, at higher energies than those with 
``conventional'' softening.

\end{itemize}               

\begin{acknowledgements}

\mbox{This work has been supported by the Swiss Science Foundation.}  

\end{acknowledgements}

\appendix

\end{document}